\documentclass[fleqn,twoside]{article}
\usepackage{espcrc2}

\usepackage{graphicx}
\usepackage[figuresright]{rotating}

\newcommand{\AmS}{{\protect\the\textfont2
  A\kern-.1667em\lower.5ex\hbox{M}\kern-.125emS}}

\title{ Topological Charge Correlators, 
Spectral Bounds, and Contact Terms
        \thanks{Talk presented by H. Thacker}}
\author{
        H. Thacker\address{Dept. of Physics, University of Virginia, Charlottesville, VA 22904},
        S.J.~Dong\address[UK]{Department of Physics \& Astronomy,
        University of Kentucky, Lexington, KY 40506, USA},
        T.~Draper\addressmark[UK],
        I.~Horv\'ath\addressmark[UK],
        F.X.~Lee\address{Center for Nuclear Studies,
        George Washington University, Washington, DC 20052, USA}\address[JL]
        {Jefferson Lab, 12000 Jefferson Avenue, Newport News, VA 23606, USA},
        K.F.~Liu\addressmark[UK],
        J.B.~Zhang\address{CSSM and Dept. of Physics and Math. Physics,
        University of Adelaide, Adelaide, SA 5005, Australia},
       }
       
\begin{document}

\begin{abstract}
The structure of topological charge fluctuations in the QCD vacuum is strongly restricted by the
spectral negativity of the Euclidean correlator for $x\neq 0$ and the presence of a positive
contact term. Some examples are considered which illustrate the physical origin of these properties.
\vspace{1pc}
\end{abstract}

\maketitle

\section{Introduction}

Although topological charge plays 
a fundamental role in our understanding of low-energy hadron physics, 
the detailed structure of topological charge fluctuations in the QCD 
vacuum is not well understood.
The construction of a local topological charge density operator for QCD \cite{hasenfratz} in terms of 
a Dirac operator with GW symmetry \cite{neuberger} has made it possible 
not only to study local $q(x)$ distributions in Monte Carlo generated gauge
fields, but to analyze these distributions
in terms of an eigenmode expansion for the corresponding Dirac operator. As discussed
in \cite{horvath}, the resulting ``eigenmode filtered'' densities
provide a physically meaningful way of removing short-wavelength
background fluctuations and focusing on whatever longer range structures might appear.
The necessity for some such filtering procedure is made clear by a fundamental
property of the two-point correlator in {\it Euclidean} 4-space\cite{seiler},
namely, that it must be {\it negative} at any nonzero separation,
\begin{equation}
\label{eq:bound}
G(x)\equiv \langle q(x)q(0)\rangle \leq 0 \;\;{\rm for}\;\; |x|\neq 0   
\end{equation}
This follows from reflection positivity (because $q(x)$ is reflection odd), or
equivalently, from spectral positivity in Minkowski space. In the latter derivation, the
negative sign of the Euclidean correlator arises from the fact that {\bf B} fields
remain real under Euclidean rotation but {\bf E} fields acquire a factor of $i$.
The bound (\ref{eq:bound}) places important restrictions on any realistic picture 
of topological charge in the QCD vacuum. For any nonzero separation, 
the positive contributions to the correlator 
from coherent, finite-size fluctuations of topological charge (e.g. instantons) 
must necessarily be 
overwhelmed by anti-correlated background fluctuations. Moreover, the requirement that 
the topological susceptibility $\chi_t=\int G(x)d^4x$ be positive implies that G(x) 
must include a positive contact term $\propto \delta^4(x)$ which makes the largest
contribution to the $\chi_t$ integral.  

The results of a numerical investigation of the topological
charge correlator in QCD \cite{horvath} indicate that it is very short range and consistent with being dominated, in the continuum limit, by an effective delta-function contact term. 
The fact that the topological charge correlator
in QCD is approximately a delta-function can also be inferred from numerical studies
of the quenched pseudoscalar hairpin correlator (i.e. the $\eta'$ mass 
insertion diagram)\cite{chlogs}. The measured correlator fits extremely well at all 
time separations to the dipole form 
$\propto (1+m_{\pi}\tau)\exp(-m_{\pi}\tau)+(\tau\rightarrow T-\tau)$. This implies that
the amputated diagram (which, in quenched QCD, is proportional to the topological charge
correlator) has very little $q^2$ dependence and is approximately a delta-function
in space-time.

In this talk, we will discuss some examples which illustrate how the negativity property 
(\ref{eq:bound}) is satisfied in practice, and also consider the physical origin of the
contact term. As a first example, consider the thermodynamics of a nonrelativistic 
free particle moving in one compact spatial dimension.\cite{arnold} Denoting the
spatial coordinate by $\phi$, the action is $S = \frac{1}{2}\dot{\phi}^2$. The 
partition function at inverse temperature $\beta$ is given by the Euclidean path
integral over all paths satisfying
\begin{equation}
\phi(\beta) = \phi(0) + C\nu
\end{equation}
where $C$ is the circumference of the compact dimension and $\nu$ is the winding 
number of the path. The winding number is the integral of a local topological charge 
density, $\nu = \int_{0}^{\beta}q(\tau)\,d\tau$ where 
\begin{equation}
q(\tau)=C^{-1}\dot{\phi}(\tau)
\end{equation}
There are classical n-instanton solutions which satisfy the Euclidean equation of motion,
\begin{equation}
\phi_n(\tau) = \frac{Cn}{\beta}\tau
\end{equation}
with action
\begin{equation}
S_n = \frac{C^2n^2}{2\beta^2}
\end{equation}
We decompose any path into the sum of an n-instanton solution and periodic fluctuations
around it,
\begin{equation}
\phi(\tau) = \phi_n(\tau) + \delta\phi(\tau)
\end{equation}
where $\delta\phi(\beta) = \delta\phi(0)$. Then it is easy to show that the topological
charge correlator separates into a sum over instantons + oscillators,
\begin{eqnarray}
\label{eq:correlator}
G(\tau)  \equiv  \langle \dot{\phi}(\tau)\dot{\phi}(0)\rangle =\nonumber\\
 \frac{C^2}{\beta^2}\sum_n n^2e^{-\beta S_n}/\sum_n e^{-\beta S_n} + &\langle\delta\dot{\phi}(\tau)
\delta\dot{\phi}(0)\rangle
\end{eqnarray}
The second term, coming from quantum fluctuations around the classical n-instanton solutions,
is obtained by differentiating the free propagator
\begin{equation}
\langle\delta\dot{\phi}(\tau)\delta\dot{\phi}(0)\rangle =
-\frac{1}{\beta}\frac{\partial^2}{\partial\tau^2} \sum_{q_j}
\frac{e^{-iq_j\tau}}{q_j^2+\lambda^2}
\end{equation}
where $q_j =\frac{2\pi j}{\beta}$.
Here $\lambda\rightarrow 0$ is a small infrared cutoff parameter. Thus, the oscillator
contribution is
\begin{equation}
\label{eq:oscillator}
\langle\delta\dot{\phi}(\tau)\delta\dot{\phi}(0)\rangle =
\frac{1}{\beta}\sum_{j\neq 0}e^{-iq_j\tau}=\delta(\tau)-\frac{1}{\beta}
\end{equation}
Now let's consider how the correlator (\ref{eq:correlator}) satisfies the bound 
(\ref{eq:bound}). There are two limiting cases of interest:\\
(I) Semiclassical or high temperature limit 
($\beta\rightarrow 0$ or $C\rightarrow \infty$).
In this limit, the instanton expansion converges, but the terms are exponentially
suppressed. The bound (\ref{eq:bound}) is satisfied because the negative term $-1/\beta$ from the
quantum fluctuations (\ref{eq:oscillator}) is always larger than the positive instanton
contribution. In this case, if we introduce a $\theta$ term, the instanton expansion gives
a good description of $\theta$ dependence (e.g. topological susceptibility). \\
(II) Ultra-quantum mechanical or low temperature limit 
($\beta\rightarrow \infty$ or $C\rightarrow 0$). 
In this case, the instanton sum diverges. Instead of expanding in
winding number, the instanton series may be resummed by a Poisson transformation
\begin{equation}
\sum_ne^{-n^2/\alpha} = \sqrt{\pi\alpha}\sum_m e^{-\alpha\pi^2 m^2}
\end{equation}
Using this formula, we find, in the large $\beta$ limit,
the resummed instanton expansion $\rightarrow +1/\beta$
Thus, in this limit, the $-1/\beta$ from quantum fluctuations exactly
cancels the instanton contribution, leaving only the contact term,
\begin{equation}
G(\tau) \rightarrow \delta(\tau)
\end{equation}
Note that, in case II, the expansion of the Poisson-resummed instanton
series is in no sense an expansion in number of instantons, but is in fact dual to it. 
It's convergence corresponds to a breakdown of the usual instanton expansion. In some
respects, this case may be viewed as a greatly oversimplified analog of Witten's 
picture of the QCD vacuum, in which topological susceptibility is finite, but is not
properly described in terms of an instanton expansion.

Note the origin of the contact term in this 0+1 dimensional
example. In momentum space, the two factors of $q$ coming from the derivatives in
the definition of the topological charge operators exactly cancel the
$1/q^2$ pole of the propagator. This is a manifestation of ``vacuum seizing''\cite{kogut},
originally discussed in the Schwinger model as a possible mechanism for resolving
the U(1) problem in QCD.

As a second example of a Euclidean topological charge correlator, we consider the
CP(N-1) sigma model in two space-time dimensions. We have studied the topological
charge correlator via (1) the large N expansion, (2) a lattice strong coupling
expansion, and (3) numerical Monte Carlo calculations.  A complete discussion of this study
will be presented elsewhere \cite{brelidze}. 
The results all indicate the dominance of the contact term in the TC correlator. 
First consider the large N
expansion. It is well-known that, to leading order in large N, 
the auxiliary U(1) gauge field develops a kinetic term and becomes dynamical due
to scalar loop effects, thereby generating a long range (confining) Coulomb potential. Thus the 
gauge field correlator behaves like
\begin{equation}
\int d^2x e^{iq\cdot x}\langle A_{\mu}(x)A_{\nu}(0)\rangle
\approx \frac{1}{q^2}\left(-g_{\mu\nu}+\frac{q_{\mu}q_{\nu}}{q^2}\right)
\end{equation}
The corresponding Euclidean correlator for the topological charge operator
$q(x)=\epsilon^{\mu\nu}\partial_{\mu}A_{\nu}$ thus produces a contact term,
\begin{equation}
\int d^2x e^{iq\cdot x}\langle q(x) q(0)\rangle
\approx const.
\end{equation}

We have also studied the TC correlator for CP(N-1) in a lattice strong-coupling expansion and
by Monte Carlo simulation. For the time-dependent correlator 
\begin{equation}
G(x_0) = \int dx_1 \langle q(x) q(0) \rangle|_{x_0=\tau} \; ,
\end{equation}
we find that, for CP(1) in the region $0\leq\beta\leq 2.0$ (correlation length $\leq 35$), 
$G(\tau)$ is completely dominated by a contact term of the form $G(\tau)= C_0\delta(\tau)
+C_2\delta''(\tau)$, with $C_0\rightarrow \approx 0$ in the weak coupling region.
Calculations for larger N models are in progress.

Based on these examples, one might suspect that a 
short range topological charge correlator dominated by a positive contact term is associated with
a strong-coupling vacuum structure for which a description based on classical instanton
solutions is inappropriate. In view of the recent QCD results \cite{horvath} it is interesting 
to ask whether a theoretical mechanism exists for generating such a contact term in QCD.
Using an operator product expansion, it is found that not only does 
the OPE predict the existence and approximate magnitude of the contact term \cite{bardeen},
but the calculation itself closely resembles the vacuum seizing mechanism encountered in
the simpler examples. The key point is that, because $\langle F^2\rangle\neq 0 $ in the QCD
vacuum, there is a term in the OPE for the $F\tilde{F}$ correlator where one gluon carries the
large momentum $q$, while a pair of soft gluons (one from each source) disappears into the vacuum.
Just as in the simpler examples, the
$1/q^2$ pole of this gluon propagator is cancelled by the momentum factors 
which arise from derivatives in the definition of the topological charge operator. 
A straightforward calculation
shows that this effective one-gluon exchange graph gives a contact term 
$\propto \delta^4(x)$, and also $\propto \langle F^2\rangle$. Using the QCD sum rule 
estimate of $\langle F^2\rangle$, one finds \cite{bardeen} that this contact term makes
a contribution to the $\eta'$ mass of 
$m_{\eta'} \approx \sqrt{\alpha_S}\times 400\; MeV$.

\end{document}